# Impact of surface treatments on the electron affinity of nitrogen-doped ultrananocrystalline diamond

Andre Chambers,[1,2*] Daniel J. McCloskey,[2] Nikolai Dontschuk,[2] Hassan N. Al Hashem,[3] Billy J. Murdoch,[4] Alastair Stacey,[2,5] and Steven Prawer,[2] and Arman Ahnood[2,3]

[1]Department of Mechanical Engineering, University of Melbourne, Victoria 3010, Australia

[2]School of Physics, University of Melbourne, Melbourne, Victoria 3010, Australia

[3]School of Engineering, RMIT University, Melbourne, Victoria 3000, Australia

[4]RMIT Microscopy and Microanalysis Facility, RMIT University, Melbourne, Victoria 3000, Australia

[5]School of Science, RMIT University, Melbourne, Victoria 3000, Australia

Keywords: UNCD, diamond, electron affinity, UPS, band diagram, surface treatment

ABSTRACT

*In recent years, various forms of nanocrystalline diamond (NCD) have emerged as an attractive group of diamond/graphite mixed-phase materials for a range of applications from electron emission sources to electrodes for neural interfacing. To tailor their properties for different uses, NCD surfaces can be terminated with various chemical functionalities, in particular hydrogen and oxygen, which shift the band edge positions and electron affinity values. While the band edge positions of chemically terminated single crystal diamond are well understood, the same is not true for nanocrystalline diamond, which has uncontrolled crystallographic surfaces with a variety of chemical states as well as graphitic grain boundary regions. In this work, the relative band edge positions of as-grown, hydrogen terminated, and oxygen terminated nitrogen-doped ultrananocrystalline diamond (N-UNCD) are determined using ultraviolet photoelectron spectroscopy (UPS), while the band bending is investigated using photoelectrochemical measurements. In contrast to the widely reported negative electrode affinity of hydrogen terminated single crystal diamond, our work demonstrates that hydrogen terminated N-UNCD exhibits a positive electron affinity owing to the increased surface and bulk defect densities. These findings elucidate the marked differences in electrochemical properties of hydrogen and oxygen terminated N-UNCD, such as the dramatic changes in electrochemical capacitance.*

## 1. INTRODUCTION

Single-crystal diamond (SCD) has long been known to display remarkable changes in interfacial electronic properties depending on its chemical surface termination. This is most apparent with the significant increase in surface conductivity associated with hydrogen termination, as well as an observed negative electron affinity (NEA) [1,2]. Similarly, other terminations, such as oxygen, fluorine, and nitrogen have been shown to lead to changes in electron affinity and the introduction of surface donor or acceptor defect states [3–7]. However, the effects of surface termination on the electron affinity of other types of diamond, viz., poly-and-nanocrystalline diamond, which incorporate significant $sp^2$ bonded carbon in the grain-boundary regions, is more complex. In addition to surface termination, the properties of this form of diamond can be adjusted for different applications by tuning the composition of the grain-boundaries as well as their relative ratio to the diamond grains [8]. Furthermore, post-growth surface etching of the grain boundary regions offers a convenient method for fine-tuning the interfacial electronic properties of these materials [9,10].

---

[*] andre.chambers@unimelb.edu.au

One prominent example of such materials is nitrogen-doped ultrananocrystalline diamond (N-UNCD), which is typically comprised of 2-5 nm diamond grains surrounded by 0.2-0.3 nm $sp^2$ carbon grain boundary regions [11]. N-UNCD has a variety of applications which exploit its unique combination of chemical stability, high bulk conductivity, biocompatibility, and tuneable surface properties. Applications include electron emission devices (photo, field, and thermionic types) for use in flat panel displays [12], free electron lasers [13], and portable mass spectrometers for extraplanetary use [14]. It is also used in electrochemical applications, including biomedical technologies such as neural stimulation and recording electrodes [15–19]. For both electron emission and electrochemical uses, the high bulk conductivity of N-UNCD is advantageous for avoiding charging effects [20]. At the same time, the interfacial electronic properties such as the electron affinity and emission intensity can be adjusted through chemical surface treatments [21].

N-UNCD has recently been shown to exhibit dramatically different electrochemical properties depending on the minutiae of the chemical surface termination [9]. Oxygen and hydrogen surface functional groups produce surface dipoles with opposite polarities when terminating the diamond surface, which are known to respectively inhibit and facilitate the direct transfer of electrons across the interface [22]. However, it has also been found that electrode properties such as the electrochemical capacitance and water window depend significantly on exactly how the surface has been chemically terminated [9,23,24]. Here, we seek to elucidate the underlying reasons for this unexpected electrochemical behaviour using surface science techniques to investigate the effect of surface termination on the band alignment of N-UNCD electrodes with redox energy levels in solution. We find that changes in the band alignment associated with small changes in the surface termination can explain the observed behaviour of these surfaces. To measure the band edge positions and Fermi level of N-UNCD, a combination of vacuum and solution-based experiments were utilised. Ultraviolet photoelectron spectroscopy (UPS) was performed for oxygen treated, hydrogen treated, and untreated (as-grown) surfaces, allowing the determination of the electron affinity and work function of the different samples. Electrochemical measurements of the open circuit potential (OCP) and flat band potential (FBP) were also performed to provide an independent measure of the surface electronic structure in an aqueous environment and understand the band bending at the interface of the electrodes with solution. We find that the band edge positions are consistent with observed electrochemical and electron emission properties, such as the extraordinarily high capacitance of oxygen annealed N-UNCD [9,23], and high photoemission of hydrogen terminated N-UNCD [25,26]. These results are compared with previously reported results from single crystal, polycrystalline, and nanocrystalline diamond, discussing the significance for potential applications.

## 2. METHODS

The N-UNCD thin-film samples used in this study were grown in a microwave plasma-assisted chemical vapour deposition (CVD) system (*IPLAS GmbH*) on nanodiamond-seeded p-type conductive silicon substrates, with a resistivity of 0.80 Ω.cm. 5×5 mm films were grown to a thickness of approximately 30 μm. Details of the N-UNCD seeding and deposition processes are reported elsewhere [27]. Samples underwent further treatment to terminate the surface with either hydrogen or oxygen. Hydrogen termination was achieved by treating samples in a pure hydrogen microwave plasma in the same CVD system as used for diamond growth. The chamber pressure was held at 30 Torr while a microwave power of 1.2 kW was applied for 10 minutes. The sample stage was held at 700°C during this process. Oxygen terminated surfaces were prepared by annealing samples in oxygen ambient at 420°C for 20 hours as described previously [23]. Our earlier work has demonstrated that these surface terminations also result in the etching of grain boundaries at the surface layer [22]. The surface morphology and chemical composition of these materials has been previously characterised [9,22].

UPS measurements were collected in vacuum using an X-ray photoelectron spectrometer (*Kratos Axis Supra*) equipped with a concentric hemispherical analyser. UPS spectra were acquired using He(I) (21.2 eV) and He(II) (40.8 eV) illumination modes from a helium discharge lamp source under a bias of 9 V. The chamber pressure was maintained below $1.5 \times 10^{-7}$ mbar during data acquisition. Samples were mounted on copper holders using carbon tape. No charge neutralisation was applied and no correction of the data to account for charging effects was necessary due to the high sample conductivity.

The open circuit potential and flat band potential of the N-UNCD electrodes in solution was investigated using a three-electrode cell connected to a potentiostat (*Gamry Interface 1010E*). The electrochemical cell consisted of a custom-made chamber containing 0.15 M NaCl solution, with a Pt disk counter electrode (1 mm diameter) and Ag/AgCl reference electrode (*eDAQ*). The NaCl solution was chosen to allow direct comparison between this work and previous studies on N-UNCD electrochemistry [9,22,23], as well as to provide the redox couples thought to be active within the adsorbed water layer on the diamond surface [28]. The working electrode was the N-UNCD sample, placed underneath an opening in the bottom of the electrochemical chamber and sealed with a 3 mm diameter O-ring (see Fig. 1). A 0.7 W near-infrared (NIR) (808 nm) laser diode (*Wuhan Lilly Electronics*) was utilized as a light source in the photocurrent experiments. The diode was pulsed using a waveform generator (*RIGOL DG4062*). The light illuminating the samples had a maximum intensity of 0.41 W $mm^{-2}$ and the illuminated area was 1.7 $mm^2$.

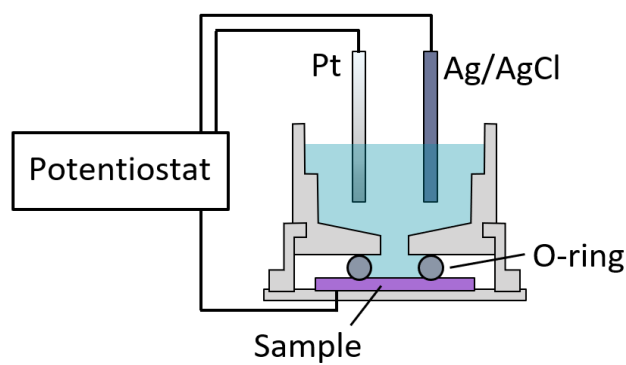

Fig. 1: Illustration of the electrochemistry chamber, indicating the placement of the working (sample), counter (Pt) and reference (Ag/AgCl) electrodes.

## 3. RESULTS

The band edge positions of the N-UNCD electrodes in solution were determined through a combination of measurements in vacuum and solution. In vacuum, UPS data allow the band edge positions relative to the vacuum energy level to be obtained. In solution, the position of the Fermi level can be ascertained by measuring the open-circuit potential (OCP). Furthermore, solution-based photocurrent analysis provides the 'flat band' potential, which in conjunction with the OCP yields information about band bending near the electrode surface [29]. The relationship between these experiments is shown schematically in the Supplementary Material (Fig. S1).

### 3.1 Ultraviolet photoelectron spectroscopy

He-I ultraviolet photoelectron spectroscopy (UPS) spectra (energy = 21.2 eV) were obtained for as-grown (AG), oxygen annealed (OA) and hydrogen plasma (HP) treated N-UNCD samples, as shown in Figure 2(a). In the He-I spectrum, the binding energy is referenced with respect to the Fermi level ($E_F$), which is calibrated using a standard gold surface and is considered the origin of the energy scale. The linear extrapolation of photoemission intensity to zero (known as $E_{CUT\text{-}OFF}$, also displayed in Figure 2(a)) can be used to determine the vacuum energy level ($E_{VAC}$) relative to $E_F$, i.e., the work function $\phi$, with the results shown in Table 1. This is achieved through the following relation:

$$\begin{aligned}\phi &= E_{VAC} - E_F \\ &= h\upsilon(\text{He-I}) + E_{CUT-OFF} \\ &= 21.2 \text{ eV} + E_{CUT-OFF}\end{aligned} \quad (1)$$

None of the samples displayed the characteristic emission shoulder and sharp increase in intensity around $E_{CUT\text{-}OFF}$ indicative of negative electron affinity (NEA), which suggests all samples exhibited positive electron affinity (PEA) [30]. HP N-UNCD displayed the lowest work function at 4.2 eV, while the work functions of AG and OA N-UNCD were approximately the same at 4.8 eV and 4.9 eV, respectively.

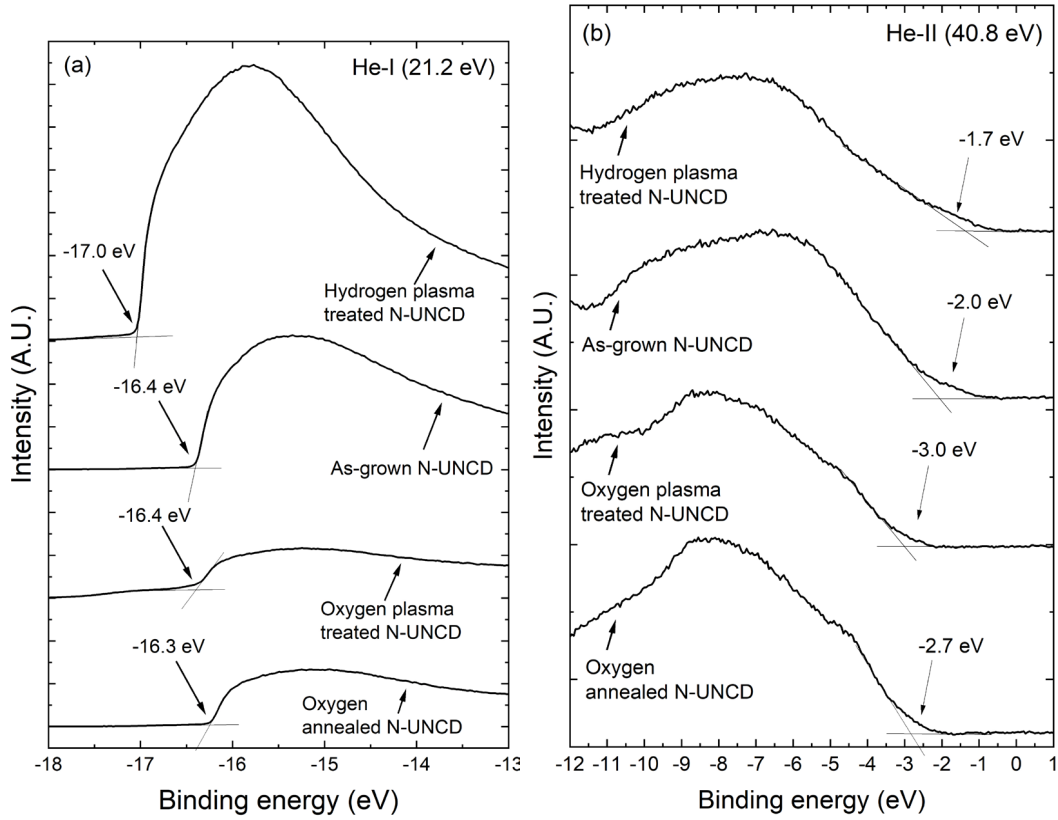

Fig. 2. (a) The low kinetic energy part of the He-I (hν = 21.2 eV) spectra of hydrogen plasma treated, as-grown, oxygen plasma treated, and oxygen annealed N-UNCD, showing the cut-off energy as determined by linear extrapolation of the onset of emission. (b) The He-II (hν = 40.8 eV) spectra of the same samples, showing linear extrapolations to find the valence band maximum. The binding energy is referenced relative to the Fermi level $E_F$.

He-II UPS spectra (energy = 40.8 eV) were also collected as shown in Figure 2(b). From these spectra the valence band maximum (VBM) relative to $E_F$ can be determined from linear extrapolation of the onset of emission. This allows calculation of the electron affinity $\chi$ from the equation:

$$\begin{aligned}\chi &= (E_f - E_{VBM}) - \phi - E_g \\ &= (E_f - E_{VBM}) - \phi - 5.5 \text{ eV}\end{aligned} \quad (2)$$

Where $E_{VBM}$ is the valence band maximum energy, and $E_g$ is the band gap of diamond, taken here to be 5.5 eV. It should be noted that this assumption about the band gap value for N-UNCD should be treated with some caution, since it is a mixed-phase material with small (2-5 nm) crystallite size [31,32]. Indeed, previous scanning tunnelling spectroscopy measurements of the band gap for hydrogen terminated N-UNCD suggest a value as low as 3.9 eV and even lower for oxygen terminated N-UNCD [33]. Nevertheless, we maintain an assumed value of 5.5 eV to be consistent with previous studies [34–36]. All quantities based on this assumption are indicated hereafter. It should also be noted that even if the band gap is assumed to be smaller than 5.5 eV, a positive value of $\chi$ will remain positive, and does not affect the relative change in $\chi$ between different N-UNCD samples. The obtained values of $\chi$ are shown in Table 1.

| Sample | $E_{CUT-OFF}$ (eV) | Work function $\phi$ (eV) | $E_F - E_{VBM}$ (eV) | Electron affinity $\chi$ (eV) |
|---|---|---|---|---|
| Hydrogen plasma treated N-UNCD | -17.0 | 4.2 | -1.7 | +0.4 |
| As-grown N-UNCD | -16.4 | 4.8 | -2.0 | +1.3 |
| Oxygen plasma treated N-UNCD | -16.4 | 4.8 | -3.0 | +2.3 |
| Oxygen annealed N-UNCD | -16.3 | 4.9 | -2.7 | +2.1 |

Table 1. The cut-off energy ($E_{CUT-OFF}$) of the UPS He-I spectra and calculated work function $\phi$, as well as the relative VBM position from the He-II spectra and calculated electron affinity $\chi$ for each N-UNCD sample.

All samples investigated show positive electron affinity, with HP N-UNCD displaying the lowest value at +0.4 eV, while AG N-UNCD and OA N-UNCD exhibit electron affinities of +1.3 eV and +2.1 eV respectively.

### 3.2 Electrochemical measurements

Further information about the electronic band structure of the N-UNCD samples was obtained from open circuit potential ($E_{OCP}$) and flat band potential ($E_{FB}$) measurements in an aqueous environment. Open circuit potential measurements allow the positions of the Fermi level to be determined for each N-UNCD surface, which are summarised in Table 2 (raw data shown in the Supplementary Information). Furthermore, solution-based flat band potential measurements were obtained by pulsed photocurrent transient analysis following previous works [37,38]. The difference between $E_{OCP}$ and $E_{FB}$ is illustrated in Figure 3.

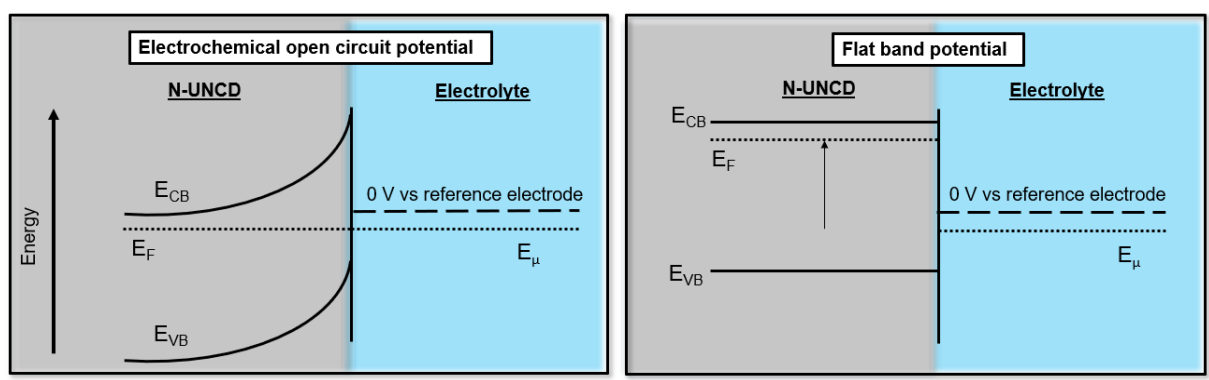

Figure 3. Simplified band diagrams showing the difference between (a) the open circuit (equilibrium) potential ($E_{OCP}$), and (b) the flat band potential ($E_{FB}$) in solution, showing the position of the conduction band minimum ($E_{CB}$), valence band maximum ($E_{VB}$), Fermi level ($E_F$), electrochemical potential of redox species in solution ($E_\mu$), and the reference electrode potential. In ideal systems, the potential difference between $E_{OCP}$ and $E_{FB}$ is equivalent to the magnitude of the band bending, as discussed further below.

As shown in Figure 4, photocurrent transients were recorded for each N-UNCD sample in response to pulsed 808 nm illumination across different applied voltages. Depending on the applied voltage, the photocurrent displayed a combination of Faradaic (direct transfer of charge, leading to monophasic waveforms) and capacitive (indirect transfer of charge, leading to biphasic waveforms) as shown in previous works [22,23]. The flat band potential can be identified as the applied voltage for which the positive and negative photocurrent transients are equal in magnitude [37]. As previously described, the sub-band gap photoresponse to 808 nm (1.5 eV) light is due to the high density of defect states relating to the grain boundary regions [22,23].

As shown in Fig. 4, the flat band potentials as determined by photocurrent transients were +0.1 V, +0.3 V, -0.1 V, -0.1 V versus the Ag/AgCl reference electrode for AG, HP, OP, and OA N-UNCD, respectively. Flat band potential values can be converted to the vacuum energy scale by subtracting the absolute potential of the Ag/AgCl reference electrode of -4.6 eV as shown in Table 2 [39]. OCP measurements for each sample were also undertaken as detailed in the Supplementary Information, with the measured values also summarised in Table 2. Theoretically, the difference between $E_{FB}$ and $E_{OCP}$ is equivalent to the magnitude of band bending at the electrode interface [29,37]. Based on this, the band bending at the AG, HP, OP, and OA N-UNCD interfaces can be inferred to be +0.2 eV, +0.3 eV, +0.3 eV, and +0.1 eV, respectively.

Qualitatively, the photocurrent transients display some marked differences between the hydrogen terminated samples and oxygen terminated samples. The AG and HP N-UNCD samples display large negative photocurrent transients at negative applied voltages, indicative of Faradaic charge transfer at the interface with solution [22–24]. On the other hand, the OP and OA N-UNCD samples exhibit positive photocurrent transients suggestive of a capacitive charge transfer mechanism [22–24]. The reasons for these differences are discussed further in Section 3.3.

It should be noted that Mott-Schottky measurements were also undertaken to validate the flat band potential values. However, these results were confounded by large double-layer capacitance values as described in the Supplementary Information. Nevertheless, the Mott-Schottky measurements do provide information about energy distribution of surface states, as also detailed in Table 2. They also show that around the open circuit potential, the AG and HP N-UNCD samples exhibit p-type conductivity, while the OP and OA samples exhibit n-type conductivity. This is likely due to the effect of surface transfer doping of the hydrogen terminated samples [28].

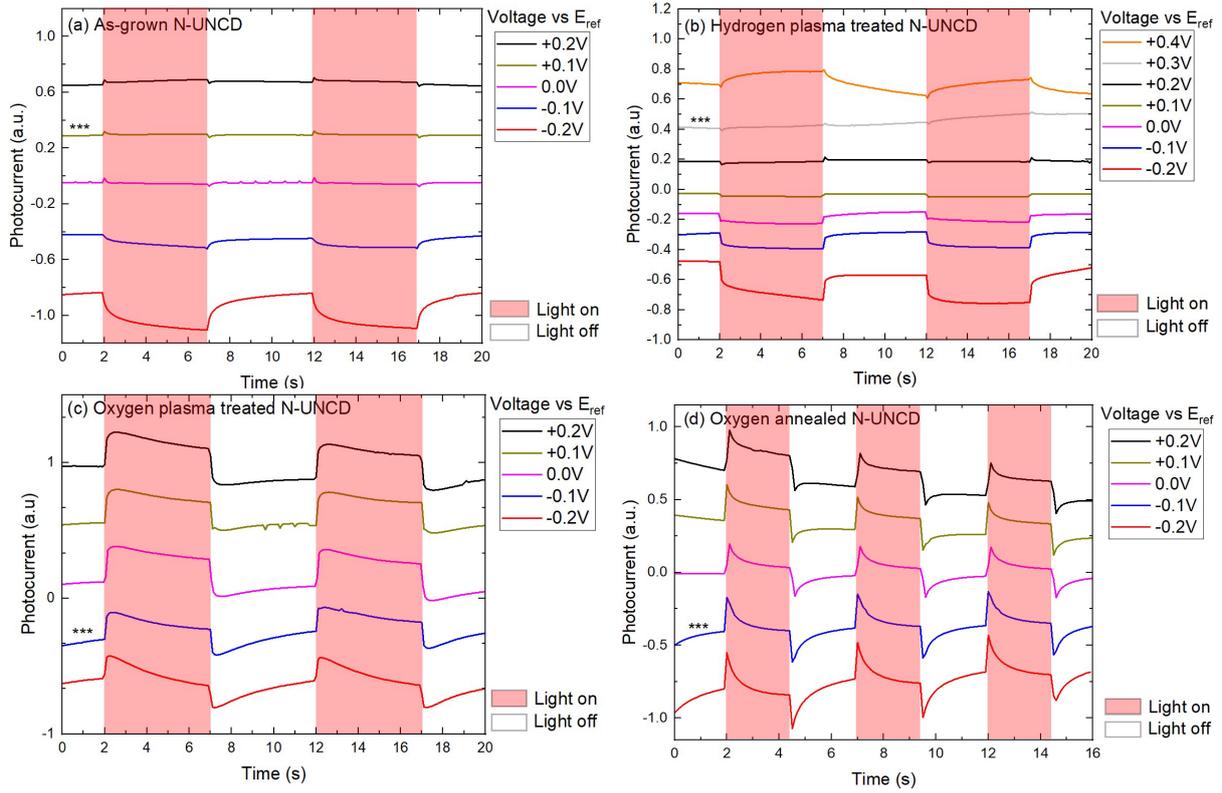

Figure 4. Flat band potential determination by photocurrent transient measurements. The photocurrent response of (a) as-grown, (b) hydrogen plasma treated, (c) oxygen plasma treated, and (d) oxygen annealed N-UNCD to 808 nm pulsed illumination with 0.7W optical power. The flat band potential is identified by the voltage applied versus the Ag/AgCl reference electrode that produces equal positive and negative photocurrent transients, as indicated by the *** symbols.

| Material | Open circuit potential vs Ag/AgCl (V) | Open circuit potential on vacuum energy scale (eV) | Flat band potential $E_{FB}$ vs Ag/AgCl (V) | Flat band potential $E_{FB}$ on vacuum energy scale (eV) | Surface states energy levels (eV) |
|---|---|---|---|---|---|
| Hydrogen plasma treated N-UNCD | 0.0 | -4.6 | +0.3 | -4.9 | - |
| As-grown N-UNCD | +0.2 | -4.8 | +0.1 | -4.7 | - |
| Oxygen plasma treated N-UNCD | +0.2 | -4.8 | -0.1 | -4.5 | -4.9 to -5.2 |
| Oxygen annealed N-UNCD | 0.0 | -4.6 | -0.1 | -4.5 | -4.9 to -5.2 |

Table 2. The open circuit potential, flat band potential and surface state energies from electrochemical measurements relative to the Ag/AgCl reference electrode ($E_{ref}$), and on the vacuum energy scale.

### 3.3 Proposed band energy diagram

N-UNCD is a mixed phase material, comprising 2-5 nm diamond grains surrounded by graphitic grain boundary regions [11]. This means that the band structure significantly differs to single crystal diamond, particularly with the inclusion of $\pi$ and $\pi^*$ sub-bands originating from $sp^2$ states in the grain boundaries [40]. As shown in Figure 5, the Gaussian shaped sub-bands are separated by approximately 2.1 eV [32]. Nitrogen defects are also expected to be found predominantly within the grain boundary regions, which here we take to be 1.7 eV below the conduction band minimum as a simplifying assumption [41]. However, it has been shown previously that hydrogen and oxygen treatments etch away the graphitic content from the surface, leaving a layer of chemically terminated diamond [22].

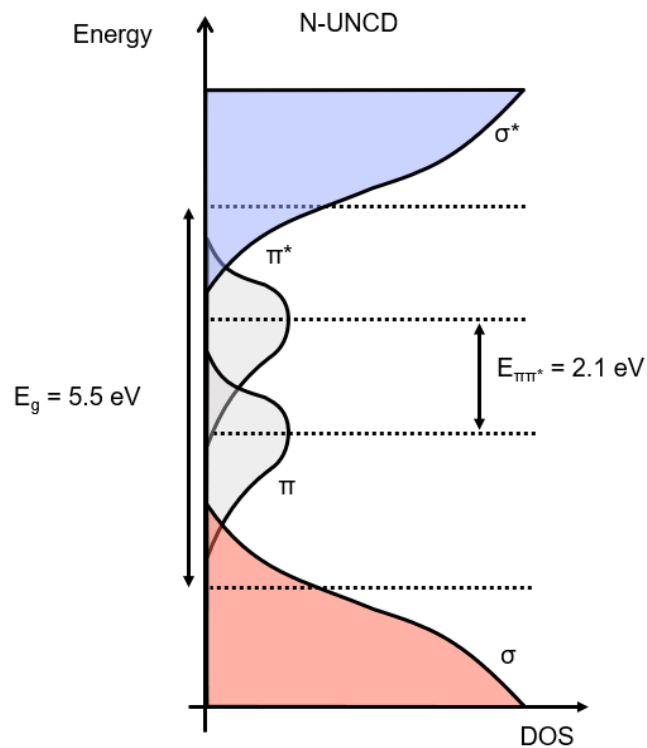

Fig. 5. A schematic band diagram of N-UNCD, showing the $\pi$ and $\pi^*$ sub-bands related to the graphitic grain boundaries. The density of states (DOS) of the sub-bands near the surface are expected to significantly reduce after the grain boundaries are etched due to hydrogen and oxygen surface treatments.

The work function and electron affinity results allow determination of the band edges relative to the vacuum energy level. The resulting band diagrams of the AG, HP, and OA N-UNCD samples are illustrated in Figure 6, showing the relative redox potentials of oxygen ($O_2/H_2O$) and hydrogen ($H^+/H_2$) related electrochemical reactions in aqueous solution at pH=7 [42]. The reduction potential for reactive carbon associated with the grain boundaries is also displayed, as this reaction has previously been observed for as-grown N-UNCD, but not for hydrogen or oxygen treated surfaces [22,43] owing to the absence of graphitic grain boundaries at the surface. The Fermi levels as determined by UPS and OCP analysis are both displayed for comparison.

It is proposed that both oxygen and hydrogen terminated N-UNCD exhibit upward band bending as suggested by the electrochemical analysis, and in agreement with previous work on single crystal diamond [44]. In particular, the HP N-UNCD valence band maximum approaches the electrochemical potential of the $O_2/H_2O$ redox couple, which may explain the greater observed Faradaic reactivity as observed in solution-based photocurrent experiments here and previously [23], as well as the p-type conductivity shown by Mott-Schottky measurements which may be due to surface transfer doping (see Supplementary Information). The Fermi levels of the AG and HP samples as determined by UPS and OCP measurements fall between the $H^+/H_2$ and $O_2/H_2O$ redox couple energy levels and generally agree within 0.5 eV. The differences may stem from the higher concentration of water molecules adsorbed to the N-UNCD surface in the OCP experiment compared to UPS. In the case of the OA sample, the Mott-Schottky data suggest a high density of interface states. The energy range of these states aligns with the Fermi level indicating Fermi level pinning [45]. These surface states are consistent with previously reported oxygen defect states which are observed in many semiconductors [47].

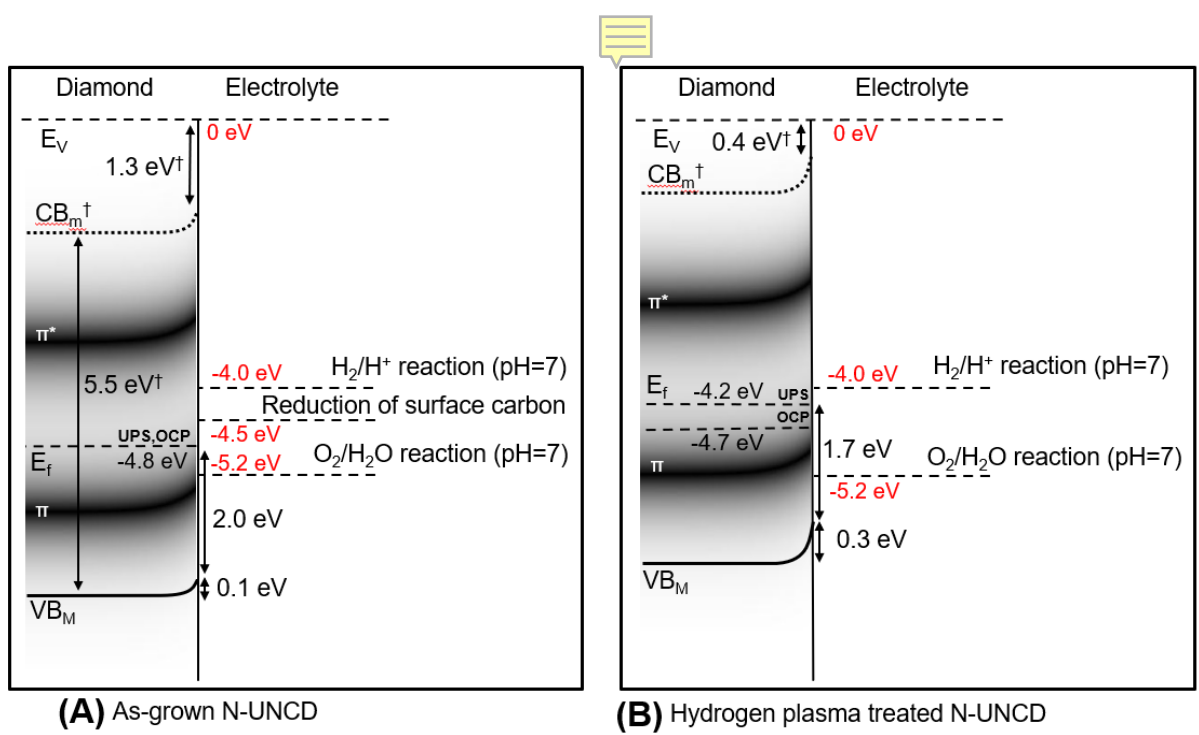

(A) As-grown N-UNCD   (B) Hydrogen plasma treated N-UNCD

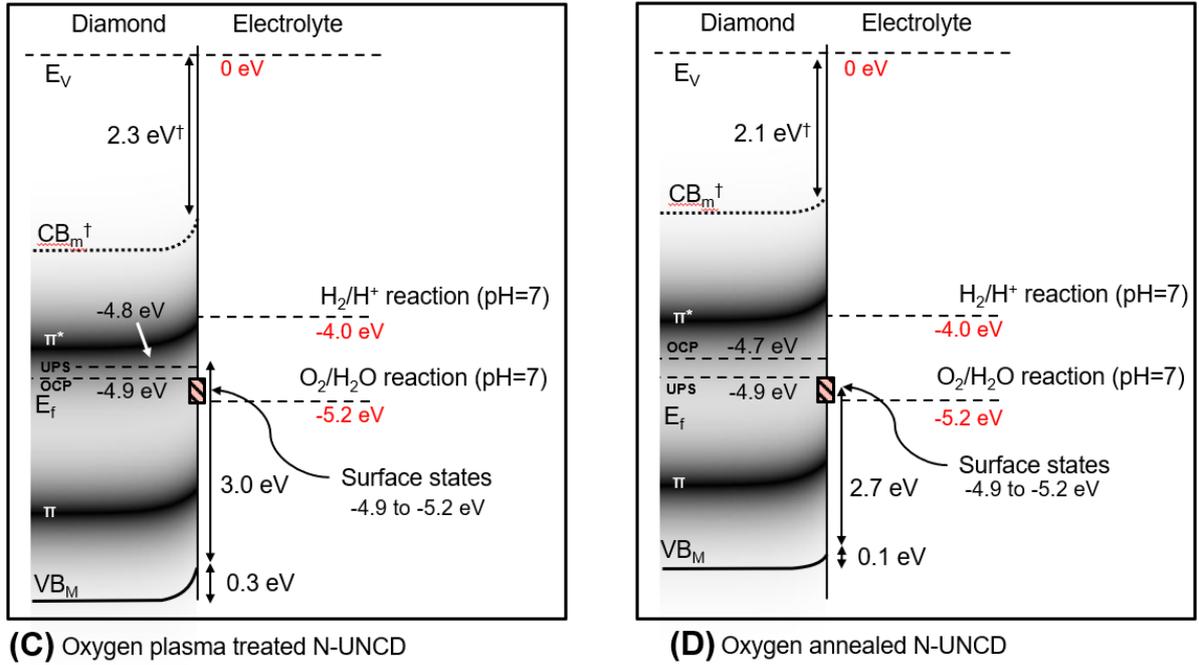

Fig. 6. Band diagrams of (a) as-grown (AG), (b) hydrogen plasma (HP) treated, (c) oxygen plasma (OP) treated, and (d) oxygen annealed (OA) N-UNCD, showing the position of the Fermi level as determined by UPS and OCP measurements relative to the electrochemical potential of redox reactions in the adsorbed water layer. The $\pi$ and $\pi^*$ sub-bands are represented by regions of shading within the band gap, with darker shading corresponding to a higher density of states. Quantities marked by † are based on the assumption of a band gap of 5.5 eV, which may not necessarily be valid for nanocrystalline diamond films as discussed in Section 3.1. Accordingly, the $CB_m$ has been tentatively marked with a dotted line.

## 4 DISCUSSION

The UPS results presented in this work are broadly consistent with previous studies on the electron emission properties of nanocrystalline diamond (NCD) and ultrananocrystalline diamond (UNCD) films [26,34–36,48,49]. However, they highlight important differences between the properties of N-UNCD and other diamond materials with larger grain sizes and lower levels of doping. Here, we find that HP treated N-UNCD exhibits a positive electron affinity (PEA) of +0.4 eV, with larger values for AG and OA samples at +1.3 eV and +2.1 eV respectively. In contrast, it is well established that hydrogen-terminated single crystal diamond displays a NEA due to the strong dipole moment of the hydrogen surface ligands [50]. NEA is not always achieved as-received after surface hydrogenation, and often requires an additional thermal anneal in order to remove the adsorbate layer and shift the EA to more negative values [51]. For example, Maier *et al.* reported an electron affinity of approximately -1.0 eV for the (100) surface as-received and -1.3 eV after being thermally cleaned to remove surface adsorbates [50]. Similarly, works studying NCD have reported NEA for hydrogen terminated samples ranging from -0.3 eV to -1.3 eV, and PEA for oxygen terminated samples ranging from +0.5 eV to +1.7 eV [34–36]. These results are summarised in Table 3 below.

Previous works have also commonly assumed NEA for N-UNCD, noting the increased quantum efficiency of photoemission after hydrogen termination [26,49]. While we observed a PEA for hydrogen terminated N-UNCD, this could be explained by a few factors. Firstly, we have deliberately not thermally cleaned the N-UNCD samples in order to probe the interface between N-UNCD and the adsorbate layer. The second factor is that the N-UNCD surface is highly heterogenous, with at least 10% of the as-grown surface expected to be $sp^2$ hybridised carbon at the grain boundaries, although the hydrogen plasma surface treatment is expected to etch the grain boundaries to some extent [22,52]. This contrasts with other types of diamond films, with the overall $sp^2/sp^3$ ratio for NCD, UNCD, N-UNCD having been reported as 0.97, 1.24, 1.51, respectively [53]. The heterogeneous surface may also contribute to an overall PEA in our N-UNCD samples; while individual diamond grains may exhibit NEA, this is mitigated by the PEA of the near-surface graphitic content within the UPS spot size of 110 μm. In addition, as-grown diamond films are known to have a layer of adsorbate hydrocarbons which may strongly affect the electron affinity [54]. This may partially explain the lower electron emission observed for AG N-UNCD compared to HP N-UNCD despite both being hydrogen terminated, since the plasma termination utilised for HP N-UNCD etches the graphitic grain boundaries at the surface and removes the hydrocarbon adsorbate layer [22,54].

| Reference | Type of diamond | Type of surface treatment | Electron affinity (eV) | Work function (eV) |
|---|---|---|---|---|
| [50] | Undoped (100) SCD | Hydrogen plasma | -1.0 (as-received) -1.3 (thermally cleaned) | 3.4 (as-received) 3.6 (thermally cleaned) |
| [36] | Undoped NCD | Hydrogen plasma | -0.3* | 3.9 |
| [35] | Undoped NCD | Hydrogen plasma | -0.1 to -0.4 (as-received)* -1.3 to -1.6 (thermally cleaned) | 4.6 (as-received) 3.4-3.8 (thermally cleaned) |
| [34] | Undoped NCD | Hydrogen plasma | -0.7 to -1.3* | |
| [26] | N-UNCD | Hydrogen plasma | | 3.0-3.1 |
| [49] | N-UNCD | Hydrogen plasma | | 4.3 |
| This work | N-UNCD | Hydrogen plasma | +0.4 (as-received)* | 4.2 |
| [50] | Undoped (100) SCD | Electrochemical oxidation | +1.7 | 6.3 |
| [36] | Undoped NCD | UV treatment | +1.6* | 4.5 |
| [35] | Undoped NCD | UV treatment | +0.7 to +1.7* | 4.6-5.1 |
| [34] | Undoped NCD | Oxygen plasma, piranha oxidation, UV treatment | +1.6, +1.6, +1.8* | |
| [48] | Undoped UNCD | UV treatment | | 4.1-4.8 |
| This work | N-UNCD | Oxygen annealed | +2.1* | 4.9 |

Table 3. A summary of electron affinity and work function values of diamond from past works. SCD, NCD, UNCD, and N-UNCD denote single crystal diamond, nanocrystalline diamond, ultrananocrystalline diamond, and nitrogen-doped ultrananocrystalline diamond, respectively. Quantities marked with * are based on the assumption of a band gap of 5.5 eV, which is not necessarily valid for nanocrystalline diamond films as discussed in Section 3.1.

Regarding the work function of N-UNCD, we obtained results of 4.2 eV, 4.8 eV, and 4.9 eV for HP, AG, and OA N-UNCD respectively. This is consistent with similar previous studies, although values range significantly. For example, Kim *et al.* measured a work function of 4.1 eV to 4.8 eV for UV treated UNCD depending on the treatment time [48], while Quintero *et al.* measured 3.0-3.1 eV for H-terminated N-UNCD and 3.6 eV for AG N-UNCD – with both works using photoemission data [26]. On the other hand, Chen *et al.* measured a work function of 4.3 eV for H-terminated N-UNCD using a Kelvin probe [49]. These differences are likely to be related to the various methods of the work function measurement, as well as variations in the doping concentration and growth conditions of the diamond films.

In terms of band line up relative to redox energy levels in solution, past work on nanodiamonds (NDs) proposed that surface transfer doping after hydrogen termination occurs due to the valence band edge overlapping with the oxygen redox couple at approximately -5.2 eV on the vacuum energy scale, while after oxygen termination the valence band is approximately 1 eV lower [28]. Here, we predict that the valence band of HP N-UNCD lies at approximately -5.9 eV. However, this is an average value for the mixed-phase N-UNCD surface, and we expect that diamond grains have a different surface potential to the grain boundary regions. This is an important distinction between NDs and N-UNCD: NDs do not possess grain boundary regions and so the surface properties can be treated similarly to those of a SCD surface.

Nevertheless, since the proximity of the valence band maximum to the redox energy level will have a significant impact on the redox activity of the electrode, this band edge position may explain the highly tuneable electrochemical properties of N-UNCD, where the HP surface facilitates Faradaic charge transfer and the OA surface displays capacitive behaviour, as seen in the photocurrent measurements presented in Section 3.2 [9,23]. This Faradaic charge transfer may be mediated by surface states as observed previously for diamond surfaces [55]. Furthermore, the opposite polarity of the photocurrent transients between hydrogen and oxygen terminated N-UNCD may be explained by the opposing electric dipole moments of the C-H and C-O surface bonds [22].

Regarding the Fermi level of each surface, we found all samples have Fermi levels approximately in the middle of the band gap, from -4.7 eV to -4.9 eV as obtained from open circuit potential measurements. This may be attributed to surface transfer doping causing pinning to the oxygen electrochemical reaction, as has been observed across many semiconductors [47]. Oxygen-related surface states are evidenced here by a plateau in the Mott-Schottky plot of OA N-UNCD between +0.3 V and +0.6 V versus the Ag/AgCl reference electrode (-4.9 eV and -5.2 eV on the vacuum energy scale), consistent with the oxygen bulk defect level at ~-5.0 eV [47].

These results have important implications for scientific and industrial applications of N-UNCD, such as electron emission and biomedical technologies. To explain the mechanism of electron emission, multiple works claim that photoemission is correlated with the density of grain boundary regions [26,49]. However, here we see that the greatest photoemission is from the HP treated sample, which we previously showed etches the graphitic grain boundary content from the surface [22]. We believe that it is important to distinguish between AG and HP N-UNCD when discussing electron emission. For AG N-UNCD, there is indeed evidence of a high proportion of emission from the grain boundary regions [13,22,56]. However, other works show nearly uniform photoemission and thermionic emission across the surface for H-terminated N-UNCD [57]. This is supported by the work of Harniman *et al.*, who suggest that electron emission is mainly from grain boundaries when the diamond grains are not conducting, but when the diamond grains are conducting (either by surface transfer doping or high doping of impurities) electron emission is spread between the grains and grain boundaries [10]. Therefore, we suggest that the enhancement of emission of HP N-UNCD may be due to a combination of complete hydrogen termination of the diamond grains and decreasing the proportion of grain boundaries at the surface, shifting the electron affinity to lower values. These findings may assist in future applications of N-UNCD electrodes. For example, the high electron emission of HP N-UNCD in solution may be readily adapted for photocatalysis and phototherapeutic applications which require sources of solvated electrons [58].

For biological applications, the reduced reactivity and increased capacitance of the OA N-UNCD, in combination with high biocompatibility, is favourable for neural stimulation and recording applications [9,16]. This high capacitance was partly attributed to etching of the grain boundary regions and the energetic separation of the valence band edge relative to redox energy levels [9]. Here, we have confirmed this energetic separation, measuring the valence band edge of OA N-UNCD at approximately 2.5 eV from the energy level of the oxygen redox couple, in comparison to approximately 0.7 eV for HP N-UNCD. It may be possible to further optimise the neural stimulation/recording properties of N-UNCD by identifying chemical functional groups that can increase this separation further.

## 5   SUMMARY AND CONCLUSIONS

This work has elucidated the significant effect of chemical surface treatment on the electronic band structure of N-UNCD, determining the work function and electron affinity of oxygen and hydrogen terminated surfaces. While previous works have determined these quantities for diamond films with larger grain sizes, the electronic properties of N-UNCD are distinct in that they are notably affected by graphitic content in the film. As shown by UPS and electrochemical measurements, AG and particularly HP N-UNCD displayed band edges shifted higher than OP and OA N-UNCD, with corresponding reductions in positive electron affinity and work function. The electron affinities of HP, AG, OP, and OA N-UNCD were determined to be +0.4 eV, +1.3 eV, +2.3 eV, and +2.1 eV, respectively. Notably, the valence band maximum of hydrogen terminated N-UNCD samples were significantly closer to the oxygen redox couple than that of the oxygen terminated samples, which may contribute to its increased electrochemical reactivity. These results assist in explaining the diverse electrochemical properties of N-UNCD depending on chemical surface termination, including the Faradaic photocurrent of AG and HP N-UNCD, compared to the capacitive photocurrent of OP and OA N-UNCD. These results have important implications for the future development of N-UNCD in electron emission and biomedical applications, which rely on chemical surface termination for tailoring the material properties.

## ACKNOWLEDGEMENTS

The authors acknowledge the facilities of the RMIT University's Microscopy and Microanalysis Facility (RMMF). AC is supported by an Australian Government Research Training Program (RTP) Scholarship.

STATEMENT OF INTERESTS

SP is a director and shareholder of Carbon Cybernetics, a company developing a diamond and carbon fibre based neural implant. AA is a shareholder in BrainConnect Pty Ltd, an Australian start-up developing physiological and neurophysiological and interventional solutions for a range of neurological disorders.